\documentclass[aps,prx,floatfix,showpacs,amsmath,twocolumn,amssymb,graphicx,longbibliography]{revtex4-2}
\usepackage{bm}
\usepackage{graphicx}
\usepackage{epstopdf}
\usepackage{wrapfig}
\usepackage{array} 
\usepackage{listings}
\usepackage[para,online,flushleft]{threeparttablex}
\usepackage{booktabs,dcolumn}
\usepackage{color}
 \usepackage{textpos}
\usepackage{booktabs}
\usepackage{multirow,bigdelim}
\usepackage{float}
\usepackage{upgreek} 
\usepackage[utf8]{inputenc}
\usepackage{hyperref}
\hypersetup{breaklinks=true,colorlinks=true,linkcolor=blue,citecolor=blue,filecolor=magenta,urlcolor=blue}
\usepackage{xcolor}
\usepackage{mathrsfs}

\newcommand{\pf}{\text{pf}}

\begin{document}

\title{Neural-network quantum states for ultra-cold Fermi gases}

\author{Jane Kim}
\affiliation{Department of Physics and Astronomy and Facility for Rare Isotope Beams, Michigan State University, East Lansing, Michigan 48824, USA}

\author{Bryce Fore}
\affiliation{Physics Division, Argonne National Laboratory, Argonne, Illinois 60439, USA}

\author{Giuseppe Carleo}
\affiliation{Institute of Physics, École Polytechnique Fédérale de Lausanne (EPFL), CH-1015 Lausanne, Switzerland}
\affiliation{
Center for Quantum Science and Engineering, \'{E}cole Polytechnique F\'{e}d\'{e}rale de Lausanne (EPFL), CH-1015 Lausanne, Switzerland
}

\author{Stefano Gandolfi}
\affiliation{Theory Division T-2, Los Alamos National Laboratory, Los Alamos, New Mexico 87545, USA}

\author{Morten Hjorth-Jensen}
\affiliation{Department of Physics and Astronomy and Facility for Rare Isotope Beams, Michigan State University, East Lansing, Michigan 48824, USA}
\affiliation{Department of Physics and Center for Computing in Science Education, University of Oslo, N-0316 Oslo, Norway}

\author{Alessandro Lovato}
\affiliation{Physics Division, Argonne National Laboratory, Argonne, Illinois 60439, USA}
\affiliation{Computational Science Division, Argonne National Laboratory, Argonne, Illinois 60439, USA}
\affiliation{INFN-TIFPA Trento Institute of Fundamental Physics and Applications, 38123 Trento, Italy}

\author{Jannes Nys}
\affiliation{Institute of Physics, École Polytechnique Fédérale de Lausanne (EPFL), CH-1015 Lausanne, Switzerland}
\affiliation{
Center for Quantum Science and Engineering, \'{E}cole Polytechnique F\'{e}d\'{e}rale de Lausanne (EPFL), CH-1015 Lausanne, Switzerland
}

\author{Gabriel Pescia}
\affiliation{Institute of Physics, École Polytechnique Fédérale de Lausanne (EPFL), CH-1015 Lausanne, Switzerland}
\affiliation{
Center for Quantum Science and Engineering, \'{E}cole Polytechnique F\'{e}d\'{e}rale de Lausanne (EPFL), CH-1015 Lausanne, Switzerland
}

\begin{abstract}

Ultra-cold Fermi gases display diverse quantum mechanical properties, including the transition from a fermionic superfluid BCS state to a bosonic superfluid BEC state, which can be probed experimentally with high precision. However, the theoretical description of these properties is challenging due to the onset of strong pairing correlations and the non-perturbative nature of the interaction among the constituent particles. This work introduces a novel Pfaffian-Jastrow neural-network quantum state that includes backflow transformation based on message-passing architecture to efficiently encode pairing, and other quantum mechanical correlations. Our approach offers substantial improvements over comparable ansätze constructed within the Slater-Jastrow framework and outperforms state-of-the-art diffusion Monte Carlo methods, as indicated by our lower ground-state energies. We observe the emergence of strong pairing correlations through the opposite-spin pair distribution functions. Moreover, we demonstrate that transfer learning stabilizes and accelerates the training of the neural-network wave function, enabling the exploration of the BCS-BEC crossover region near unitarity. Our findings suggest that neural-network quantum states provide a promising strategy for studying ultra-cold Fermi gases.

\end{abstract}

\maketitle

\section{Introduction}

The study of ultra-cold Fermi gases has received considerable experimental and theoretical attention in recent years due to their unique properties and potential applications in fields ranging from condensed matter physics to astrophysics. These systems can be created and manipulated in the laboratory with high precision, providing a versatile platform for investigating a wide variety of phenomena. By tuning the $s$-wave scattering length $a$ via external magnetic fields near a Feshbach resonance, one can smoothly crossover from a fermionic superfluid BCS state ($a < 0$) of long-range Cooper pairs to a bosonic superfluid BEC state ($a > 0$) of tightly-bound, repulsive dimers. Given their diluteness, the behavior of these systems is mainly governed by $a$ and the effective range of the potential $r_e$, with natural units provided by the Fermi momentum $k_F$ and the Fermi gas energy per particle in the thermodynamic limit $E_{FG} = \frac{3}{5} \frac{\hbar^2}{2m} k_F^2$ (see Ref.~\cite{Gandolfi_2014} and references therein).

The region between the BCS and BEC states, known as the ``unitary limit,'' is particularly interesting as $a$ diverges and $r_e$ approaches zero. The unitary Fermi gas (UFG) is a strongly-interacting system that exhibits surprisingly stable superfluid behavior. Studying the BCS-BEC crossover near the unitary limit can reveal critical aspects of the underlying mechanism behind superfluidity in fermionic matter. The UFG is also universal, meaning its properties are independent of the details of the two-body potential. This universality allows for robust comparisons and predictions between seemingly disparate quantum systems. For instance, the UFG is relevant for neutron stars, as they provide a means to study superfluid low-density neutron matter~\cite{Gezerlis:2008, Gandolfi:2015}, whose properties are crucial for the phenomenology of glitches~\cite{Monrozeau:2007xu} and the cooling of these stars via neutrino emission~\cite{Yakovlev:2004iq,Page:2010aw,Ho:2015}. 

The onset of strong pairing correlations and the non-perturbative nature of the interaction makes the theoretical study of these systems particularly challenging for quantum many-body methods. Among them, quantum Monte Carlo (QMC) has proven to be exceptionally efficient in calculating various properties with high accuracy, including the energy~\cite{Carlson:2011}, pairing gap~\cite{Carlson:2008}, and other quantities related to the so-called contact parameter~\cite{Gandolfi:2011}. Diffusion Monte Carlo (DMC), in particular, is an accurate tool for calculating the properties of quantum many-body systems~\cite{Foulkes:2001}. The fixed-node approximation typically employed in DMC calculations to control the fermion-sign problem provides a rigorous upper bound to the ground-state energy that agrees well with other methods and experiments~\cite{Forbes:2011,Ku:2012}. Moreover, unlike the Auxiliary Field Quantum Monte Carlo method, DMC can handle broad classes of local interactions, which provides exact results that are sign-problem free but is limited only to unpolarized systems with a purely attractive interaction~\cite{Carlson:2011}. However, the fixed-node approximation limits the accuracy of DMC energies, which induces a residual dependence on the starting variational wave function. The latter has a critical role in DMC calculations of expectation values of operators that do not commute with the Hamiltonian, such as spatial and momentum distributions. The analytical form of the variational ansatz is usually tailored to specific problems of interest and biased by the physical intuition of the researchers.

In this work, we overcome these limitations by performing variational Monte Carlo (VMC) calculations of ultra-cold Fermi gases with neural-network quantum states (NQS)~\cite{carleo_solving_2017} that incorporate only the most essential symmetries and boundary conditions.  After their initial application to quantum-chemistry problems~\cite{Hermann:2019,Pfau2020}, continuous-space NQS have been successfully employed to study quantum many-body systems in the presence of spatial periodicities, such as interacting quantum gases of bosons~\cite{Pescia:2022}, the homogeneous electron gas~\cite{Wilson2022,Cassella2023}, and dilute neutron matter~\cite{Fore2022}. Recent works have also used NQS to solve the nuclear Schr\"odinger equation in both real space\cite{Keeble:2019bkv,Adams2021,Gnech2021,Lovato2022,Yang2022} and the occupation number formalism~\cite{Rigo2022}. When dealing with fermions, the antisymmetry is usually enforced using generalized Slater determinants, the expressivity of which can be augmented with either backflow transformations~\cite{Luo2019} or by adding ``hidden'' degrees of freedom~\cite{Moreno2022}. 

Strong pairing correlations in fermionic systems motivate adopting an antisymmetrized wave function constructed from pairing orbitals rather than single-particle orbitals. One such construction, often called the geminal wave function~\cite{Casula:2003,Casula:2004}, considers determinants of spin-singlet pairs, while other more general wave functions based on the Pfaffian~\cite{Bajdich:2006,Bajdich:2008,Gandolfi:2009,Gandolfi:2020pbj}, consider both singlet and triplet contributions. Pfaffian wave functions combined with neural-network Jastrow correlators~\cite{nomura2017restricted, nomura2021dirac} have successfully modeled lattice fermions, even revealing the existence of a quantum spin liquid phase in the J1-J2 models on two-dimensional lattices.

We propose a novel NQS that extends the conventional Pfaffian-Jastrow~\cite{Bajdich:2008} ansatz by incorporating neural backflow transformations into a fully trainable pairing orbital. The backflow transformations are generated by a message-passing architecture recently introduced to model the homogeneous electron gas~\cite{Pescia:2023}. In addition to being a significant departure from generalized Slater determinants, our Pfaffian-Jastrow NQS naturally encodes pairing in the singlet and triplet channels, without stipulating a particular form for the pairing orbital. In view of this, it is broadly applicable to other strongly-interacting systems with the same symmetries and boundary conditions. We demonstrate the representative power of our NQS by computing ground-state properties of ultra-cold Fermi gases in the BCS-BEC crossover. Our Pfaffian-Jastrow NQS outperforms Slater-Jastrow NQS by a large margin, even when generalized backflow transformations are included in the latter. Most notably, we find lower energies than those obtained with state-of-the-art DMC methods, which start from highly-accurate BCS-like trial wave functions.

The rest of the paper is organized as follows. In Section~\ref{sec:methods}, we introduce the Hamiltonian used to model ultra-cold atomic gases near the unitary limit and the many-body techniques used to solve the Schr\"odinger equation. In Section~\ref{sec:results}, we compare the Pfaffian-Jastrow NQS with other NQS ans\"atze and state-of-the-art DMC results. Finally, in Section~\ref{sec:conclusion}, we draw our conclusion and provide future perspectives of this work.

\section{Methods}
\label{sec:methods}
As customary in QMC approaches, we simulate the infinite system using a finite number of fermions $N$ in a cubic simulation cell with side length $L$, equipped with periodic boundary conditions (PBCs) in all $d=3$ spatial dimensions. We use $\bm{r}_i \in \mathbb{R}^d$ and $s_i \in \{ \uparrow, \downarrow \}$ to denote the positions and spin projections on the $z$-axis of the $i$-th particle, and the length $L$ can be determined from the uniform density of the system $N / L^3 = k_F^3/(3 \pi^2)$. The dynamics of the unpolarized gas is governed by the non-relativistic Hamiltonian
\begin{equation}
H=-\frac{\hbar^2}{2m}\sum_i^N \nabla_i^2+\sum_{ij}^N v_{ij} \,,
\label{eq:ham}
\end{equation}
where the attractive two-body interaction
\begin{equation}
v_{ij} = (\delta_{s_i, s_j} - 1) v_0  \frac{2\hbar^2}{m} \frac{\mu^2}{\cosh^2(\mu r_{ij})} \,,
\label{eq:int}
\end{equation}
acts only between opposite-spin pairs, making the interaction mainly in $s$-wave for small values of $r_e$. In the above equations, $\nabla_i^2$ is the Laplacian with respect to $\bm{r}_i$ and $r_{ij} = \| \bm{r}_i - \bm{r}_j\|$ is the Euclidean distance between particles $i$ and $j$. The P\"oschl-Teller interaction potential of Eq.~\eqref{eq:int} provides an analytic solution of the two-body problem and has been employed in several previous QMC calculations~\cite{Chang:2004,Gezerlis:2009,Morris:2010,Gandolfi:2011}. The parameters $v_0$ and $\mu$ tune the scattering length $a$ and effective range $r_e$, respectively. In the unitary limit $|a| \rightarrow \infty$, the zero-energy ground state between two particles corresponds to $v_0=1$ and $r_e = 2/\mu$. In order to analyze the crossover between the BCS and BEC phases, we will use different combinations of $v_0$ and $\mu$ that correspond to the same effective range. In addition, we will consider various values of $\mu$ with fixed $v_0=1$ to extrapolate the zero effective range behavior at unitarity.

\subsection{Neural-network quantum states}
\label{subsec:nqs}
We solve the Schr\"odinger equation associated with the Hamiltonian of Eq.~\eqref{eq:ham} using two different families of NQS. 
All ans\"atze have the general form 
\begin{equation}
\Psi (X) = e^{J(X)} \Phi(X),
\label{eq:gen-form}
\end{equation}
where the Jastrow correlator $J(X)$ is symmetric under particle exchange and $\Phi(X)$ is antisymmetric. Here, we have introduced $X=\{ \bm{x}_1, \dots, \bm{x}_N \} $, with $\bm{x}_i = (\bm{r}_i, s_i)$, to represent the set of all single-particle positions and spins compactly.

In addition to the antisymmetry of fermionic wave functions, the periodic boundary conditions, and the translational symmetry (which will be discussed in Sec.~\ref{subsubsec:mpnn}), we also enforce the discrete parity and time-reversal symmetries as prescribed in Ref.~\cite{Lovato2022}. More specifically, we carry out the VMC calculations for the unpolarized gas using $\Psi^{PT}(R,S)$ given by
\begin{align}
\Psi^P(R,S) &= \Psi(R,S) + \Psi(-R,S), \\
\Psi^{PT}(R,S) &= \Psi^P(R,S) + (-1)^{n}\Psi^P(R,-S),
\end{align}
where $n = N/2$ and we have used the notation $R = \{ \bm{r}_1, \dots, \bm{r}_N \}$ and $S=\{  s_1, \dots, s_N\}$ for the set of all positions and spins, respectively. Enforcing these symmetries has been shown to accelerate the convergence of ground-state energies for both atomic nuclei~\cite{Lovato2022} and dilute neutron matter~\cite{Fore2022}.

\subsubsection{Pfaffian-Jastrow}
\label{subsubsec:pj}

The antisymmetric part of the wave function employed in QMC studies of ultra-cold Fermi gases is typically constructed as an antisymmetrized product of BCS spin-singlet pairs~\cite{Carlson:2003zz,Chang:2004,Gezerlis:2008,Gezerlis:2009,Galea:2016}. It goes by a variety of names, such as the geminal wave function~\cite{Casula:2003,Casula:2004}, the singlet pairing wave function~\cite{Bajdich:2008}, and the (number-projected) BCS wave function~\cite{Galea:2016}, just to name a few. 
Although geminal wave functions have demonstrated significant improvements over single-determinant wave functions of single-particle orbitals, the energy gains are typically smaller for partially spin-polarized systems~\cite{Casula:2004}, as contributions from the spin-triplet channel are missing.
This naturally leads to the singlet-triplet-unpaired (STU) Pfaffian wave function~\cite{Bajdich:2006,Bajdich:2008}, in which the pairing orbitals are explicitly decomposed into singlet and triplet channels. Then, the STU ansatz is expressed as the Pfaffian of a block matrix, with the singlet, triplet, and unpaired contributions partitioned into separate blocks. When the triplet blocks are zero, the STU wave function reduces to the geminal wave function. 

Both the geminal and the STU wave functions rely on fixing the spin ordering of the interacting fermions. Consequently, they are not amenable to potentials that exchange spin, such as those used to model the interaction among nucleons~\cite{Piarulli:2019cqu}. In neutron-matter calculations, for instance, the pairing orbital for the Pfaffian wave function can be taken as a product of a radial part and a spin-singlet part~\cite{Gandolfi:2009,Gandolfi:2020pbj}. The spin-triplet pairing has so far been neglected in neutron-matter calculations, but they can be treated similarly without requiring spin ordering.


To address the limitations of the works mentioned above, we take the most general form of the Pfaffian wave function~\cite{Bajdich:2006,Bajdich:2008} as the antisymmetric part of our ansatz
\begin{equation}
\Phi_{PJ}(X) = \pf
\begin{bmatrix}
  0 & \phi(\bm{x}_1, \bm{x}_2) & \cdots & \phi(\bm{x}_1, \bm{x}_N) \\
  \phi(\bm{x}_2, \bm{x}_1) & 0 & \cdots & \phi(\bm{x}_2, \bm{x}_N) \\  
   \vdots & \vdots & \ddots & \vdots \\
 \phi(\bm{x}_N, \bm{x}_1) & \phi(\bm{x}_N, \bm{x}_2) & \cdots & 0 \\
  \end{bmatrix}
  ,
  \label{eq:gen_pf}
\end{equation}
where we assume the unpolarized case for this initial investigation. We do not keep the spins fixed, nor do we mandate a specific form for the pairing orbital $\phi(\bm{x}_i, \bm{x}_j)$. Instead, we capitalize on the universal approximation property of feed-forward neural networks (FNN)~\cite{Hornik:1991} by defining the pairing orbital as
\begin{equation}
\phi(\bm{x}_i, \bm{x}_j) 
= \nu(\bm{x}_i, \bm{x}_j) 
- \nu(\bm{x}_j, \bm{x}_i),
\label{eq:pair-orb}
\end{equation}
where $\nu$ is a dense FNN. The above expression ensures that the Pfaffian is mathematically well-defined, as the matrix is skew-symmetric by construction $\phi(\bm{x}_i, \bm{x}_j) = -\phi(\bm{x}_j, \bm{x}_i)$. Since $\nu$ takes all the degrees of freedom of a given pair of particles as input, including the spins, our pairing orbital has the capacity to discover the spin-singlet and spin-triplet correlations on its own. 

This design leaves our Pfaffian-Jastrow (PJ) ansatz agnostic to any particular form of the interaction and systematically improvable by simply increasing the size of $\nu$. The input dimension of $\nu$ only depends on the spatial dimension $d$ and not the total number of particles $N$, leading to an exceptionally scalable ansatz. Given the generality of our formulation, the Pfaffian ansatz calculation cannot be reduced to a determinant of singlet pairing orbitals as in the geminal wave function. Thus, the efficient computation of the Pfaffian is crucial to the scalability of our approach. To this aim, we implement the Pfaffian computation according to Ref.~\cite{Wimmer:2012}. 


We further improve the nodal structure of our PJ ansatz through backflow (BF) transformations~\cite{feynman1956backflow}. To our knowledge, this is the first time neural BF transformations have been used in a Pfaffian wave function, although they have demonstrated their superiority over traditional BF transformations within the Slater-Jastrow formalism in numerous applications~\cite{Luo2019, Hermann:2019, Pfau2020}. We replace the original single-particle coordinates $\bm{x}_i$ by new ones $\tilde{\bm{x}}_i(X)$, such that correlations generated by the presence of all particles are incorporated into the pairing orbital. To ensure that the Pfaffian remains antisymmetric, the backflow transformation must be permutation equivariant with respect to the original $\bm{x}_i$, i.e. $\tilde{\bm{x}}_i$ depends on $\bm{x}_i$ and is invariant with respect to the set $\{\bm{x}_j\}_{j \neq i}$. In Sec.~\ref{subsubsec:mpnn}, we discuss in detail how the backflow correlations are encoded via a permutation-equivariant message-passing neural network. All calculations labeled as PJ-BF assume that we apply the transformation $\nu(\bm{x}_i, \bm{x}_j) \to \nu(\tilde{\bm{x}}_i, \tilde{\bm{x}}_j)$ to the FNN in Eq.~\eqref{eq:pair-orb}.

\begin{figure*}[tbh]
 \includegraphics[width=\textwidth]{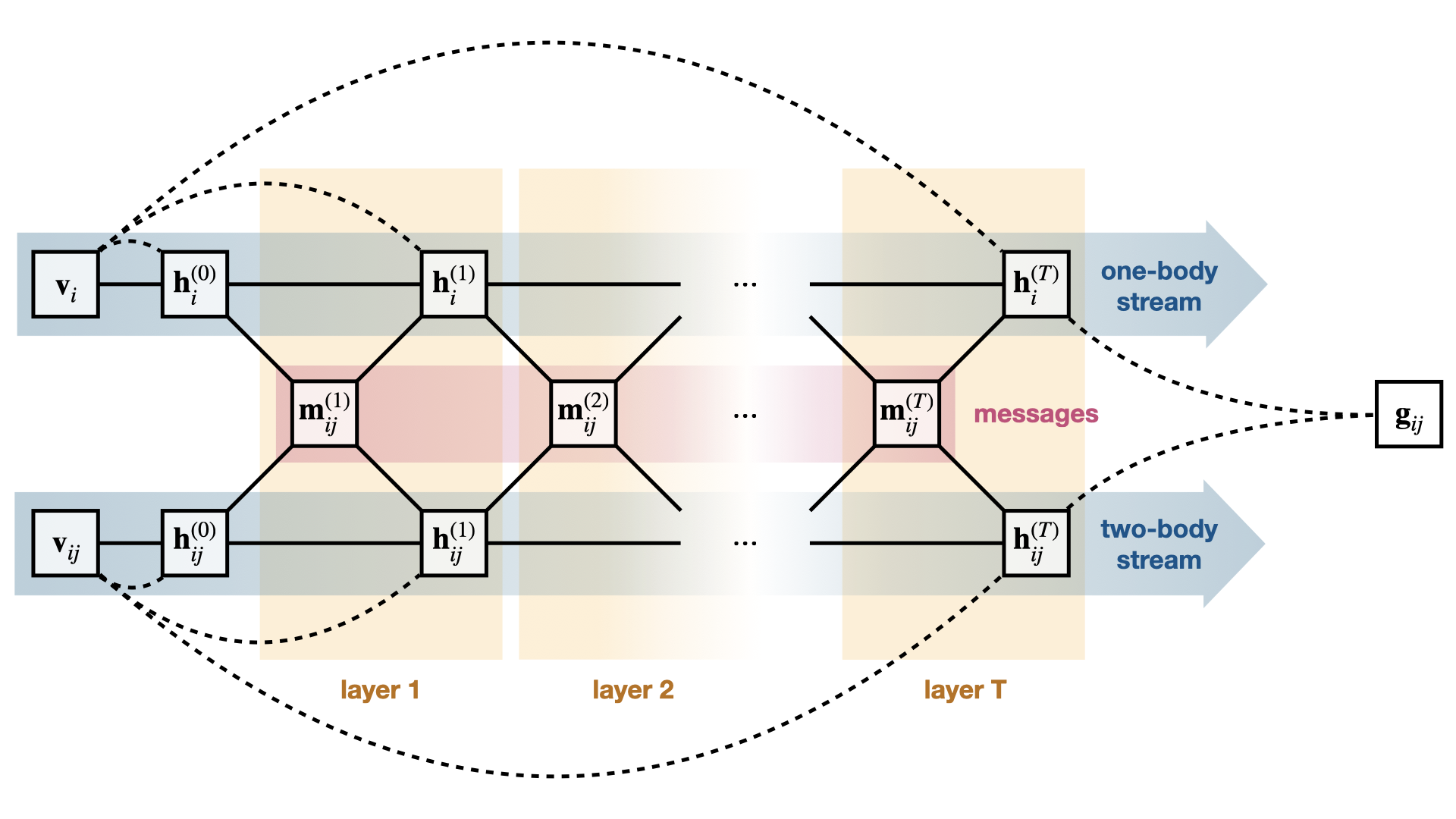}
  \caption{Schematic representation of a message-passing neural network with $T$ iterations. Dashed lines represent the concatenation operations, while solid lines represent the parameterized transformations (linear transformations and nonlinear feedforward neural networks). Messages, highlighted in pink, mediate the exchange of information between the one- and two-body streams, in blue. A yellow box indicates a single iteration of the network. }
\label{fig:mpnn}
\end{figure*}

\subsubsection{Slater-Jastrow}
\label{subsubsec:sj}
For comparison, we will report results obtained using a Slater-Jastrow (SJ) ansatz, which amounts to taking the antisymmetric part of the wave function to be a Slater determinant of single-particle states
\begin{equation}
\Phi_{SJ}(X) = \det
\begin{bmatrix}
  \phi_1(\bm{x}_1) & \phi_1(\bm{x}_2) & \cdots & \phi_1(\bm{x}_N) \\
  \phi_2(\bm{x}_1) & \phi_2(\bm{x}_2) & \cdots & \phi_2(\bm{x}_N) \\  
   \vdots & \vdots & \ddots & \vdots \\
 \phi_N(\bm{x}_1) & \phi_N(\bm{x}_2) & \cdots & \phi_N(\bm{x}_N) \\
  \end{bmatrix}
  .
\label{eq:sj}
\end{equation}
In the fixed-node approximation, the single-particle states are the products of spin eigenstates with definite spin projection on the $z$-axis $s_\alpha$ and plane wave (PW) orbitals with discrete momenta $\bm{k}_\alpha=2\pi \bm{n}_\alpha /L$,   $\bm{n}_\alpha \in\mathbb{Z}^d$,
\begin{equation}
\phi_\alpha(\bm{x}_i) = {\rm e}^{i\bm{k}_\alpha \cdot\bm{r}_i} \chi_\alpha (s_i) \, ,
\label{eq:pw-phi}
\end{equation}
where $\chi_\alpha(s_i)=\delta_{s_\alpha, s_i}$. Here, $\alpha = (\bm{k}_\alpha, s_\alpha)$ denotes the quantum numbers characterizing the state. We will label Slater-Jastrow NQS calculations using above plane wave orbitals as SJ-PW. 

As in the Pfaffian case, we improve the nodal structure of the above Slater determinant using backflow transformations generated by the message-passing neural network discussed in Sec.~\ref{subsubsec:mpnn}. We modify the spatial coordinates of Eq.~\eqref{eq:pw-phi} as
\begin{equation}
\bm{r}_i \to  \bm{r}_i + \bm{u}_i(X)\, ,
\label{eq:bf-r}
\end{equation}
where the complex backflow displacement $\bm{u}_i(X) \in \mathbb{C}^d$ allows for changes in both the phases and amplitudes of the spatial part of the single-particle states. We also map the single-particle spinors onto the Bloch sphere as
\begin{equation}
\begin{split}
\chi_\alpha(\tilde{\bm{x}}_i) 
&=\cos \left( \frac{\theta_i(X)}{2} \right) \delta_{s_\alpha, s_i}\\
&\hspace{3mm}+ \sin \left( \frac{\theta_i(X)}{2} \right) (1-\delta_{s_\alpha, s_i}),
\end{split}
\label{eq:bf-chi}
\end{equation}
where $\theta_i(X) \in \mathbb{R}$ is the polar angle on the sphere. Both $\bm{u}_i(X)$ and $\theta_i(X)$ are permutation-equivariant functions of the original coordinates $\bm{x}_i$, the functional form of which will be discussed in Sec.~\ref{subsubsec:mpnn}. Appendix~\ref{appx:bf} includes a detailed explanation of Eq.~\eqref{eq:bf-chi}. Slater-Jastrow NQS calculations using the backflow orbitals will be labeled as SJ-BF.

\subsubsection{Message-Passing Neural Network}
\label{subsubsec:mpnn}

Implementing the aforementioned neural-network quantum states is possible using $X$ as direct inputs to the appropriate FNNs and Deep-Sets~\cite{Zaheer:2018}. Still, it is advantageous to devise new inputs that already capture a large portion of the correlations. As in Ref.~\cite{Pescia:2023}, we employ a permutation-equivariant message-passing neural network (MPNN) to iteratively build correlations into new one-body and two-body features from the original “visible” features. The visible features are chosen to be
\begin{align}
\bm{v}_i &= (s_i) \, ,\\
\bm{v}_{ij} &= \left( \bm{r}_{ij}, \| \bm{r}_{ij} \| , s_{ij} \right) ,
\end{align}
with the separation vectors $\bm{r}_{ij} = \bm{r}_i - \bm{r}_j$ and distances $\| \bm{r}_{ij} \|=r_{ij}$ replaced by their $L$-periodic surrogates
\begin{align}
\bm{r}_{ij} &\mapsto \left( \cos(2 \pi \bm{r}_{ij}/ L), \sin(2 \pi \bm{r}_{ij}/ L) \right)\, ,\\
\| \bm{r}_{ij} \| &\mapsto \| \sin( \pi \bm{r}_{ij} / L) \|,
\end{align}
and the quantity $s_{ij} \equiv 2\delta_{s_i, s_j}-1$ assigned a value of $+1$ for aligned spins and $-1$ for anti-aligned spins. Note that we have excluded explicit dependence on the particle positions $\bm{r}_i$ in the visible one-body features, thereby enforcing translational invariance in the new features. Linear transformations are applied to and concatenated with each feature to obtain the initial hidden features
\begin{align}
\bm{h}_i^{(0)} &= (\bm{v}_i, A \bm{v}_i),\\
\bm{h}_{ij}^{(0)} &= (\bm{v}_{ij}, B \bm{v}_{ij}).
\label{eq:hid-init}
\end{align}
The main purpose of the linear transformations is to preprocess the input data. Still, they also help simplify the implementation by keeping the dimension of the hidden features $\bm{h}_i^{(t)}$, $\bm{h}_{ij}^{(t)}$ constant for all $t$. In each iteration, $t = 1, \dots, T$ of the MPNN, information is exchanged between the one- and two-body streams through a so-called “message”
\begin{equation}
\bm{m}_{ij}^{(t)} = \bm{M}_t \left(\bm{h}_i^{(t-1)}, \ \bm{h}_j^{(t-1)}, \ \bm{h}_{ij}^{(t-1)} \right).
\label{eq:pair-message}
\end{equation}
For a given particle $i$, relevant messages are collected and pooled together to destroy the ordering with respect to all other particles $j\neq i$,
\begin{equation}
\bm{m}_i^{(t)} = \bm{pool} \left( \{ \bm{m}_{ij}^{(t)} \ | \ j \neq i \} \right).
\label{eq:part-message}
\end{equation}
The pooling operation $\bm{pool}$ collapses the order of the elements in the set it acts upon and produces a vector with the same dimension as an individual element. Throughout this work, we use logsumexp-pooling, the smooth variation of max-pooling.

The pairwise messages $\bm{m}_{ij}^{(t)}$ and the implied particle messages $\bm{m}_{i}^{(t)}$ are then used to update the hidden features 
\begin{align}
\bm{h}_i^{(t)} &= \left(\bm{v}_i, \ \bm{F}_t\left(\bm{h}_i^{(t-1)}, \  \bm{m}_i^{(t)} \right)\right),\\
\bm{h}_{ij}^{(t)} &= \left(\bm{v}_{ij}, \ \bm{G}_t\left(\bm{h}_{ij}^{(t-1)}, \ \bm{m}_{ij}^{(t)} \right)\right).
\label{eq:hid-update}
\end{align}
The functions $\bm{M}_t$, $\bm{F}_t$, and $\bm{G}_t$ are all unique FNNs with the same output dimension as the linear preprocessors $A$ and $B$. By incorporating concatenated skip connections to the visible features, we guarantee that the signal originating from the raw data remains discernible even as the MPNN depth $T$ increases. Finally, we combine the resulting outputs $\bm{h}_i^{(T)}$ and $\bm{h}_{ij}^{(T)}$ into pairwise feature vectors
\begin{equation}
\bm{g}_{ij} = \left(\bm{h}_i^{(T)}, \bm{h}_j^{(T)}, \bm{h}_{ij}^{(T)} \right)
\end{equation}
to feed into subsequent networks. The flow of information through the MPNN can be visualized in Fig.~\ref{fig:mpnn}. Notice how the hidden features in a given layer depend on the hidden features of the previous layer \textit{and} the original visible features.

For all our NQS, we use a Jastrow correlator based on a Deep-Set~\cite{Zaheer:2018} to enforce permutation invariance over the set of all pairwise features
\begin{equation}
J(X) = \rho \Big( \bm{pool} \Big( \{ \bm{\zeta} (\bm{g}_{ij}) \ | \ i \neq j \} \Big) \Big). 
\label{eq:jastrow}
\end{equation}
Here, $\rho$ and $\bm{\zeta}$ are FNNs, and the pooling operation is the same as in Eq.~\eqref{eq:part-message}. While many Jastrow functions are typically designed to satisfy Kato's cusp condition~\cite{Kato:1957} for specific systems, we take a different approach and allow our neural networks to learn the cusp fully. The short-range behavior of the ground state is particularly important for the UFG, so leaving our NQS completely unbiased serves as an important test for evaluating the overall capabilities of NQS. 

The Slater-Jastrow ansatz with plane wave orbitals (SJ-PW) does not require any additional neural networks beyond $\rho$ and $\bm{\zeta}$, so it establishes a baseline for the number of trainable parameters in this work. On the other hand, the backflow variables $\bm{u}_i$ and $\theta_i$ for the Slater-Jastrow ansatz with backflow orbitals (SJ-BF) are the outputs of another Deep-Set
\begin{equation}
(\text{Re}(\bm{u}_i), \text{Im}(\bm{u}_i), \theta_i) = \bm{\rho}_{bf} \Big( \bm{pool} \Big( \{ \bm{\zeta}_{bf} (\bm{g}_{ij}) \ | \ j \neq i \} \Big) \Big),
\end{equation}
which is permutation invariant with respect to all $j \neq i$ by construction. The size of $\bm{\rho}_{bf}$ and $\bm{\zeta}_{bf}$ determines the number of extra variational parameters present in the SJ-BF ansatz compared to the SJ-PW ansatz. For the PJ ansatz, the pairing orbital $\nu$ in Eq.~\eqref{eq:pair-orb} simply takes $\bm{g}_{ij}$ as input in place of $(\bm{x}_i, \bm{x}_j)$. Therefore, the number of additional variational parameters in the PJ ansatz relative to the SJ-PW ansatz is determined by the size of $\nu$. 

All of the feedforward neural networks mentioned throughout this section have at least two hidden layers with 16 nodes each. The activation function is GELU~\cite{Hendrycks:2016} and the weights/biases are initialized with glorot normal/zeros unless pretrained parameters are used. 

It is worth highlighting that the individual feedforward neural networks within our NQS are solely dependent on the spatial dimension $d$ and not the system size $N$. Therefore, even though this study focuses on benchmarking the $N=14$ case, the trained NQS can be used as starting points for larger even-$N$, unpolarized systems without requiring any modifications to the network structure. This is an example of transfer learning, a powerful strategy that involves applying knowledge gained from solving one problem to another, often more challenging problem. 


\subsubsection{Variational Monte Carlo and Training}
\label{subsec:vmc}

We train our NQS by minimizing the energy 
\begin{equation}
E(\bm{p}) \equiv \frac{\langle \Psi(\bm{p}) | H | \Psi(\bm{p}) \rangle}{\langle \Psi(\bm{p}) | \Psi(\bm{p}) \rangle}
\end{equation}
with respect to the variational parameters $\bm{p}$. To compute the energy and its gradient $\nabla_{\bm{p}} E$ using Monte Carlo integration, we sample positions $R$ and spins $S$ from $|\Psi(R,S)|^2$ in a way that preserves periodicity and total spin projection on the $z$-axis, as in Refs.~\cite{Pescia:2022, Lovato2022}. Since the ordering of the spins is not fixed, our ans\"atze can be immediately applied to any continuous-space Hamiltonian that exchange spin, such as Ref.~\cite{Schiavilla:2021}. 

A sophisticated optimization technique is critical for achieving an ansatz that is both compact and expressive. In this work, we employ the stochastic reconfiguration~\cite{Sorella2005} (SR) algorithm with regularization based on the RMSprop method, introduced in Ref.~\cite{Lovato2022}. The parameters are updated as
\begin{equation}
\bm{p} \leftarrow \bm{p} - \eta G^{-1} \nabla_{\bm{p}} E,
\end{equation}
where $\eta$ is a constant learning rate and $G$ is the quantum geometric tensor~\cite{Stokes:2020}.

Due to the strong and short-range nature of the interaction in Eq.~\eqref{eq:int}, it is likely for the optimization process to get trapped in a local minimum when initialized with random parameters. To avoid this problem, we use transfer learning by pretraining the NQS on a softer interaction ($\mu = 5$) before proceeding to harder ones ($\mu = 10, 20, 40$). Not only does this approach improve the final converged energy, but the efficiency of the optimization process overall. The training for lower values of $\mu$ can handle a more aggressive learning rate $\delta$ and fewer samples. As a general guideline, we reduce $\delta$ by a factor of 10 and double the number of samples each time the value of $\mu$ is doubled. The number of optimization steps required for training ranges from $\mathcal{O}(10^3)$ to $\mathcal{O}(10^4)$, depending on whether the neural quantum states (NQS) were pretrained or initialized with random parameters.

\subsection{Diffusion Monte Carlo}
\label{subsec:dmc}

The fixed-node DMC calculations are performed as described in Ref.~\cite{Pessoa2015}. The initial state is prepared using VMC methods with a variational wave function with the same general form as Eq.~\eqref{eq:gen-form}. Note that, within the fixed-node approximation, DMC provides a strict upperbound to the energy of the system. While DMC is a precise method, its accuracy relies on the choice of nodal surface and the quality of the preceding VMC calculation. The symmetric Jastrow factor is given by
\begin{align}
J(X) &= \sum_{ii'}^n u(r_{ii'}), \\
u(r) &= K\tanh(\mu_J r)\cosh(\gamma r)/r \,,
\end{align}
where $n=N/2$ and the unprimed and primed indicies denote the spin-up and spin-down particles, respectively.
The parameters $K$ and $\gamma$ are adjusted so that $u(d)=0$ and $u'(d)=0$, and $\mu_J$ and $d$ are variational parameters. 
Considering that the $s$-wave channel dominates the interaction, the antisymmetric part is given by the number-projected BCS wave function
\begin{equation}
\Phi_{BCS}(X) = \det
\begin{bmatrix}
  \phi(\bm{r}_{11'}) & \phi(\bm{r}_{12'}) & \cdots & \phi(\bm{r}_{1n'}) \\
  \phi(\bm{r}_{21'})  & \phi(\bm{r}_{22'}) & \cdots & \phi(\bm{r}_{2n'}) \\  
   \vdots  & \vdots  &\ddots & \vdots \\
 \phi(\bm{r}_{n1'}) & \phi(\bm{r}_{n2'}) & \cdots &\phi(\bm{r}_{nn'}) \\
  \end{bmatrix}
  ,
\label{eq:bcs}
\end{equation}
with the pairing orbitals
\begin{align}
\phi(\bm{r}) &=\tilde{\beta} (r) +
\sum_{i} a(k_i^2) e^{i\bm{k}_i \cdot \bm{r}}\,, \label{eq:dmc-pair-orb}
\\
\tilde{\beta}(r)&= \beta(r)+\beta(L-r)-2 \beta(L/2) \,,
\\
\beta (r) &= ( 1 + c b  r )\ ( 1 - e^{ - d b r })
\frac{e^{ - b r} }{dbr}~ \,.
\end{align}
The parameters $a(k^2_i)$, $b$ and $d$ are obtained by minimizing the energy, and $c$ is chosen so that the function $\beta$ has zero slope at the origin. If we instead let $\beta=0$ and restrict the sum in Eq.~\eqref{eq:dmc-pair-orb} to momentum states filled up to $k_F$, the antisymmetric part is equivalent to the Slater determinant with single-particle plane waves as in Eqs.~\eqref{eq:sj} and ~\eqref{eq:pw-phi}. Since this approach does not involve pairing, we will refer to the related DMC results as DMC-PW. Conversely, the approach that accounts for pairing will be identified as DMC-BCS.

It should be emphasized that the BCS wave function of Eq.~\eqref{eq:bcs} is a special case of the generalized Pfaffian of Eq.~\eqref{eq:gen_pf}. In fact, it can be easily shown~\cite{Bajdich:2008} that by only retaining the spin-singlet blocks, the calculation of the Pfaffian reduces to the determinant of spin-singlet block.


\section{Results}
\label{sec:results}

\begin{figure}
  \includegraphics[width=\columnwidth]{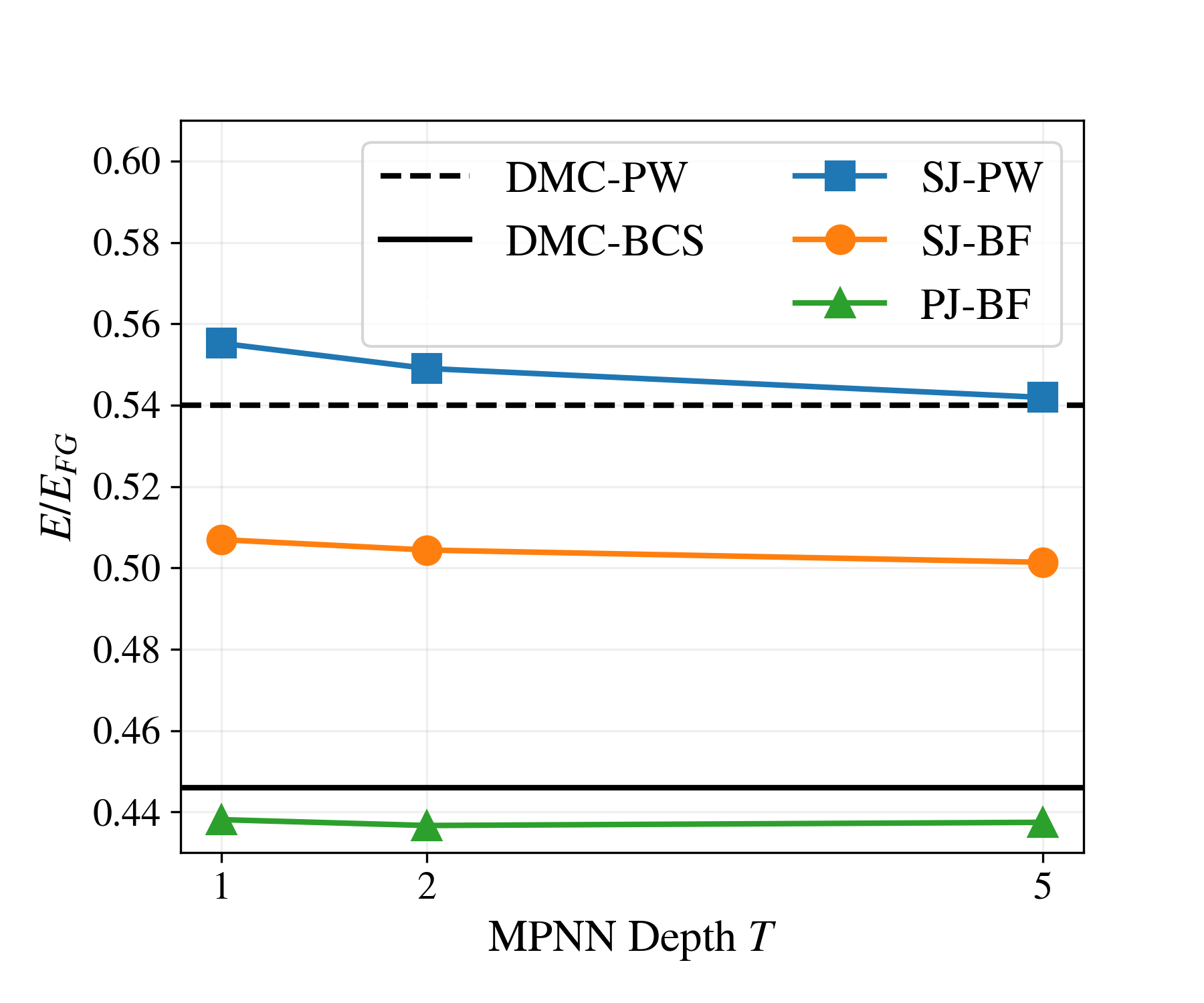}
  \caption{Ground-state energies per particle as a function of the MPNN depth $T$ for the SJ-PW (blue squares), SJ-BF (orange circles), and PJ-BF (green triangles) ans\"atze. The interaction parameters are set to $v_0 = 1$ and $\mu = 5$, corresponding to an effective range of $r_e k_F = 0.4$. The DMC benchmark energies with and without pairing are displayed as solid and dashed lines, respectively. }
\label{fig:E-T}
\end{figure}

We first compare the performance of the various neural-network quantum states outlined in Sec.~\ref{subsec:nqs} as the message-passing neural network (MPNN) depth $T$ is varied. As shown in Fig.~\ref{fig:E-T}, the final converged energies per particle for the Slater-Jastrow ansatz with plane wave orbitals (SJ-PW) decreases monotonically towards the corresponding DMC-PW benchmark, with remarkable agreement at $T=5$. This behavior echoes the findings of Ref.~\cite{Schaetzle:2021}, and demonstrates the impact of the MPNN on the flexibility of our Jastrow. Incorporating backflow correlations into the Slater-Jastrow ansatz (SJ-BF) significantly improves results compared to the fixed-node approach with PW, but more than half of the discrepancy between the two DMC energies remains. Due to the observed weak dependence on $T$, it is unlikely that further increasing $T$ would yield a substantial improvement in energy. The SJ-BF ansatz may be able to achieve energies more similar to the DMC-BCS benchmark by increasing the width of the feedforward neural networks. Still, the associated computational expenses are expected to be prohibitively high. Therefore, we turn our attention to the Pfaffian-Jastrow (PJ) ansatz. Even with a single MPNN layer, the PJ ansatz easily outperforms DMC-BCS while also possessing fewer parameters ($\sim$5600 v.s. $\sim$6200) than the single-layer SJ-BF ansatz. The overall dependence on the MPNN depth is weak, with $T=2$ giving slightly lower energy and variance than $T=5$. For the remainder of our analysis, we will use the PJ ansatz with $T=2$, which contains about 8500 variational parameters. 

\begin{figure}
\includegraphics[width=\columnwidth]{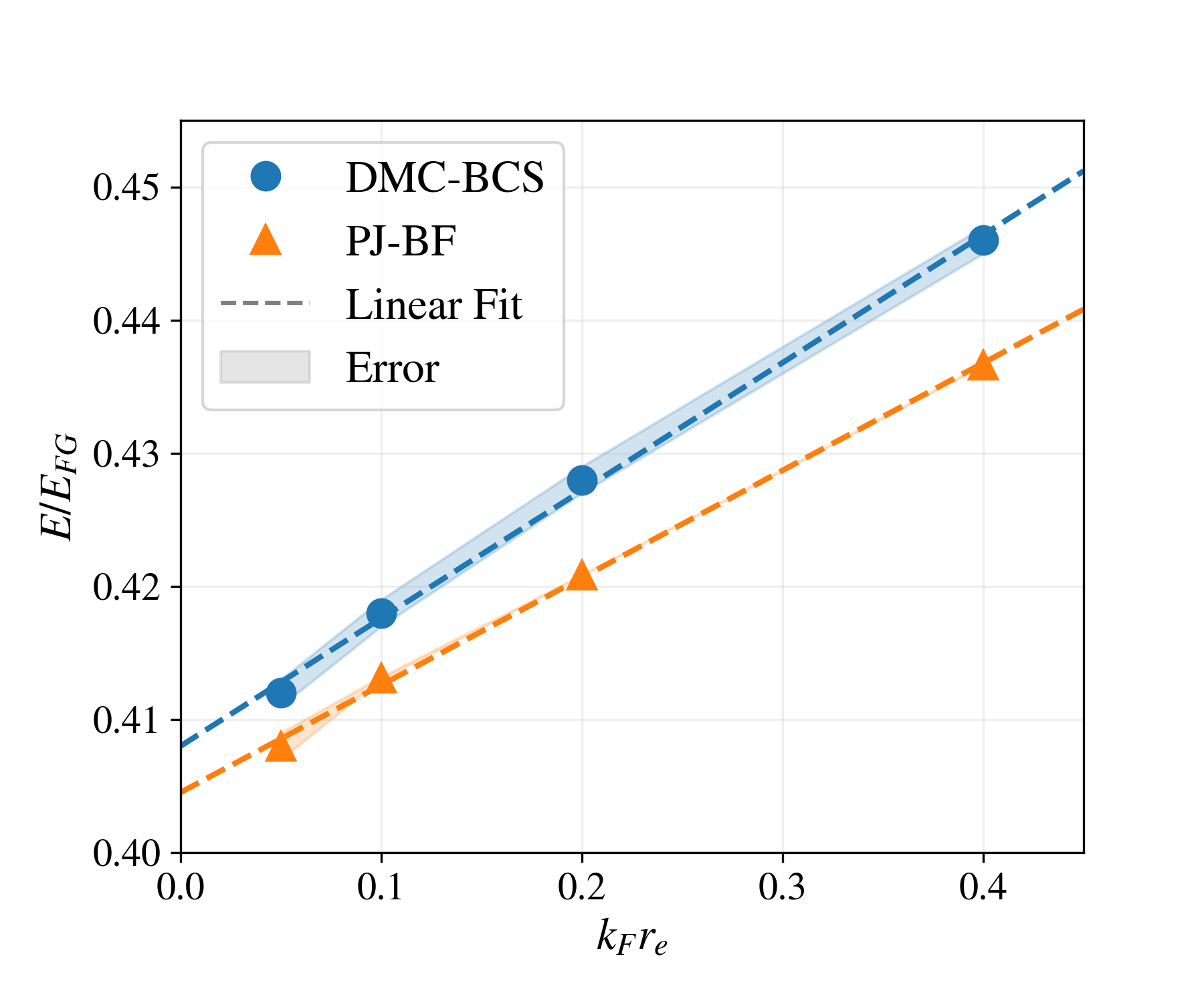}
  \caption{Ground-state energies per particle as a function of the effective range. The DMC-BCS benchmark energies (blue circles) and the Pfaffian-Jastrow with backflow (PJ-BF) energies (orange triangles) are extrapolated to zero effective range using linear fits (dashed lines). The shaded regions are the error bands for the DMC-BCS and PJ energies.}
\label{fig:E-re}
\end{figure}

\begin{table}[t]
  \centering
    \begin{tabular}{ |c|c||c|c| } 
    \hline
     $\mu$ & $k_F r_e$  & DMC-BCS & PJ-BF \\
     \hline\hline
     5 & 0.4 & 0.446(1) & 0.4366(3) \\
     10 & 0.2 & 0.428(1) & 0.4208(3) \\
     20 & 0.1 & 0.418(1) & 0.4131(8) \\
     40 & 0.05 & 0.412(1) & 0.408(1) \\
     $\infty$ & 0.0 & 0.408(1)$^*$ & 0.405(1)$^*$\\
    \hline
    \end{tabular}
\caption{Energy per particle for various values of $\mu$ and the corresponding values of $r_e$. The values with asterisks ($^*$) are extrapolations from the linear fits shown in Fig.~\ref{fig:E-re}. The parameter $v_0=1$ is fixed.}
\label{tab:E-mu}
\end{table}

As the unitary limit is characterized by a vanishing effective range, we study how the ground-state energy responds to changing $k_F r_e$ in Fig.~\ref{fig:E-re}. The PJ ansatz gives energies $\sim$1-2\% lower than DMC-BCS as the effective range is decreased from $k_Fr_e = 0.4$ to $k_Fr_e= 0.1$. At $k_F r_e = 0.05$, our energy falls below the range of the DMC-BCS error band, suggesting our approach is likely to maintain its superior performance as $r_e$ is decreased further. To estimate the energy at zero effective range, we also perform simple linear fits --- See Table~\ref{tab:E-mu} for the extrapolated values. Note that our results have been obtained by simulating a system of $N=14$ particles for benchmark purposes. In order to obtain energies closer to the thermodynamic limit, further simulations with more particles will be needed~\cite{Forbes:2011,Forbes_2012}. 

\begin{figure}
  \includegraphics[width=\columnwidth]{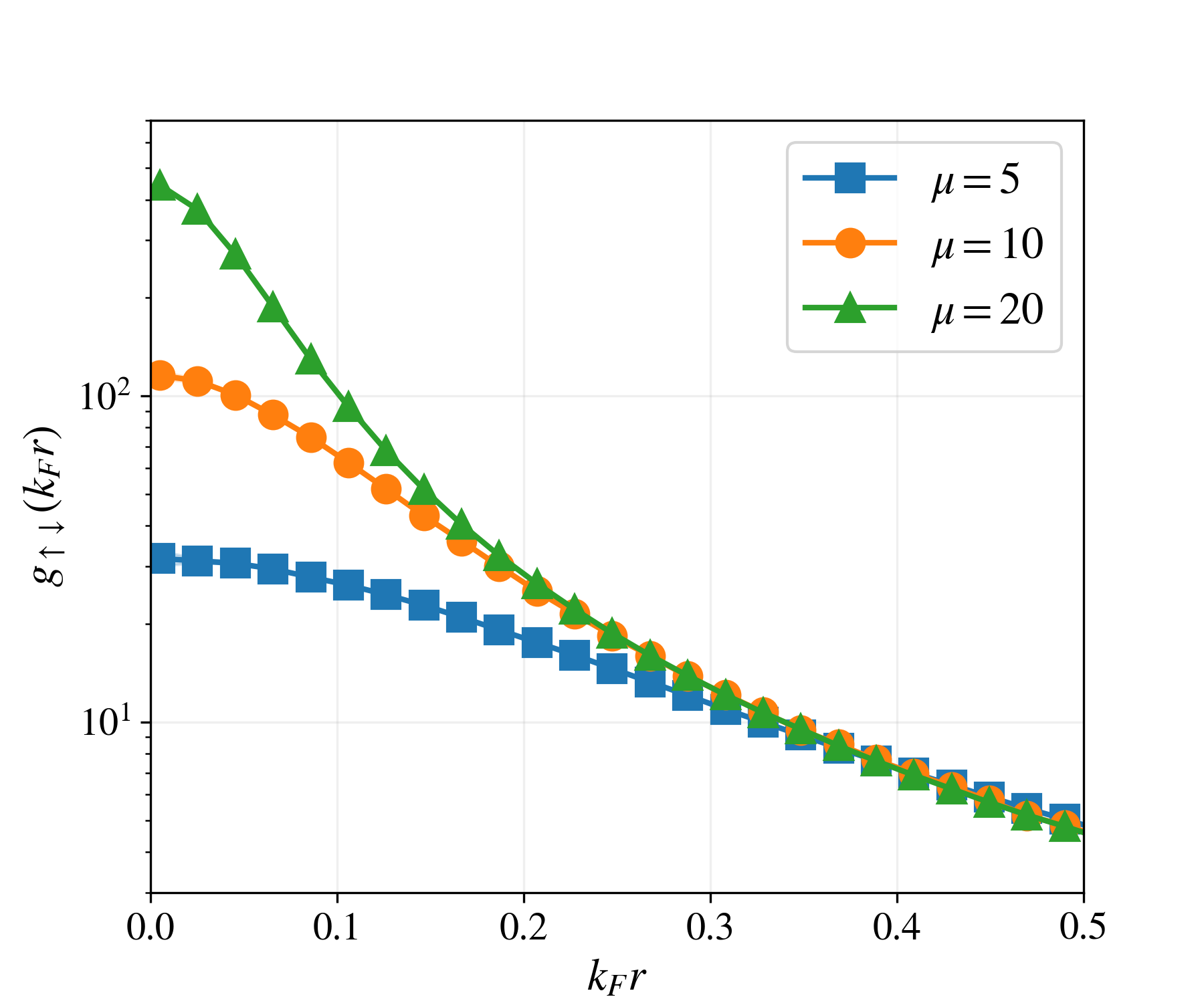}
  \caption{Opposite-spin pair densities as a function of small $k_F r$ at unitarity ($v_0 = 1$) and $\mu = 5$ (blue squares), $\mu=10$ (orange circles), and $\mu=20$ (green triangles).}
\label{fig:density-unitary}
\end{figure}

\begin{figure}
  \includegraphics[width=\columnwidth]{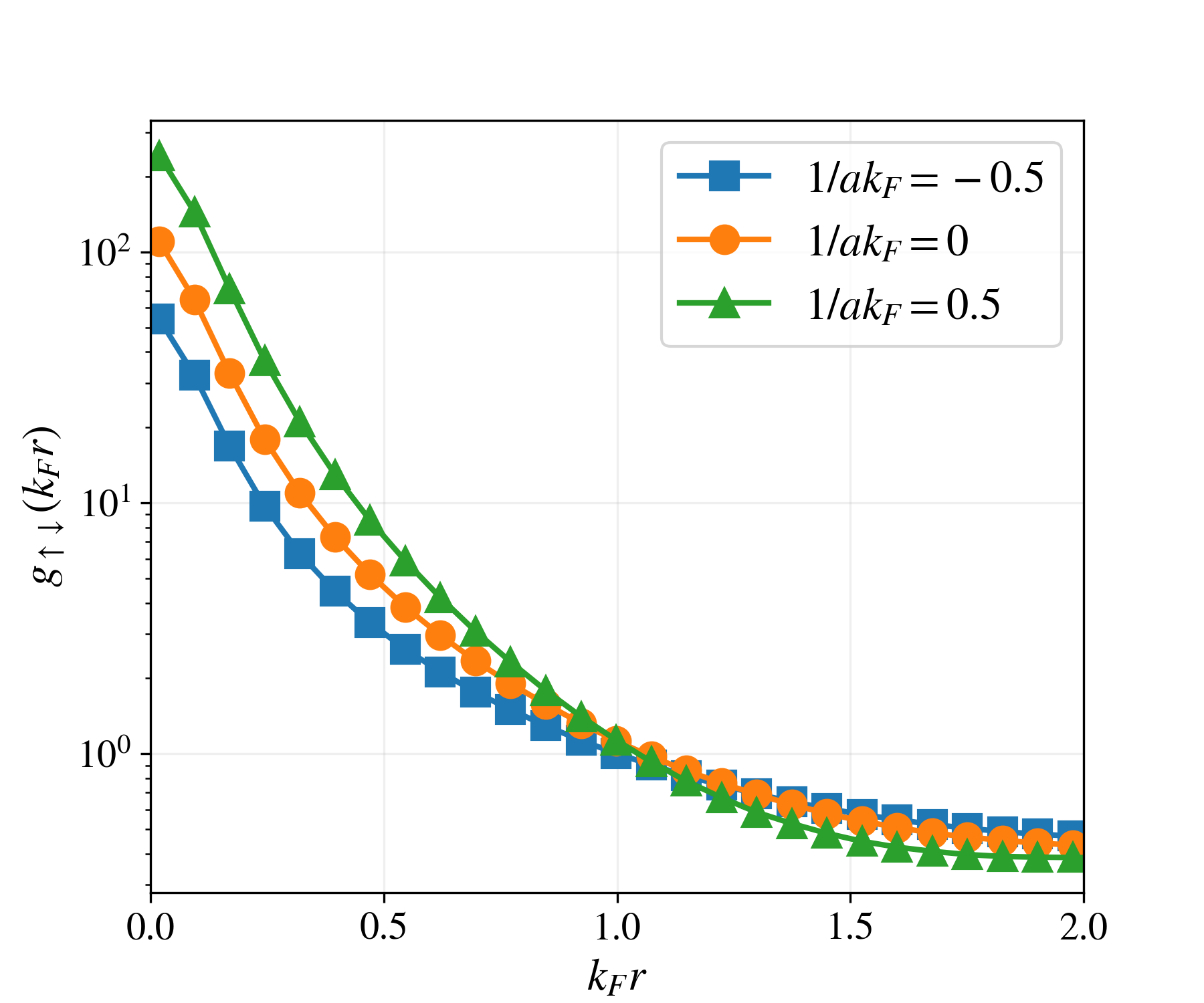}
  \caption{Opposite-spin pair densities in the crossover region for the BCS phase $1/{a k_F} = -0.5$ (blue squares), unitarity $1/{a k_F} = 0$ (orange circles), and BEC phase $1/{a k_F} = 0.5$ (green triangles). The effective range of all cases are fixed $k_F r_e = 0.2$. See Table ~\ref{tab:int-params} for the corresponding values of $v_0$ and $\mu$.}
\label{fig:density-crossover}
\end{figure}

In Fig.~\ref{fig:density-unitary}, we show the opposite-spin pair distribution functions at unitarity for $\mu = $ 5, 10, and 20. Notice how the peaks of the distributions at $k_F r = 0$ grow roughly quadratically with $\mu$, demonstrating the presence of strong pairing correlations as we approach the unitary limit $\mu \rightarrow \infty$. Clearly, the short-range character of the distributions are important to capture at unitarity, as they begin to converge around $k_F r\gtrsim 0.4$. 

Fig.~\ref{fig:density-crossover} presents a complementary set of opposite-spin pair distribution functions in the crossover region with fixed effective range of $k_Fr_e = 0.2$. When we lean towards the BCS phase $1/ak_F = -0.5$, the long-range tail of the density is enhanced compared to the unitary case $1/ak_F = 0$. On the other hand, the tail is diminished in the BEC phase $1/ak_F = -0.5$, suggesting the initiation of dimer formation. The differences in the peaks of the distributions are not as dramatic as in Fig.~\ref{fig:density-unitary}, but they are consistent with the expected behavior in the BCS and BEC regimes near unitarity. 

\begin{figure}
  \includegraphics[width=\columnwidth]{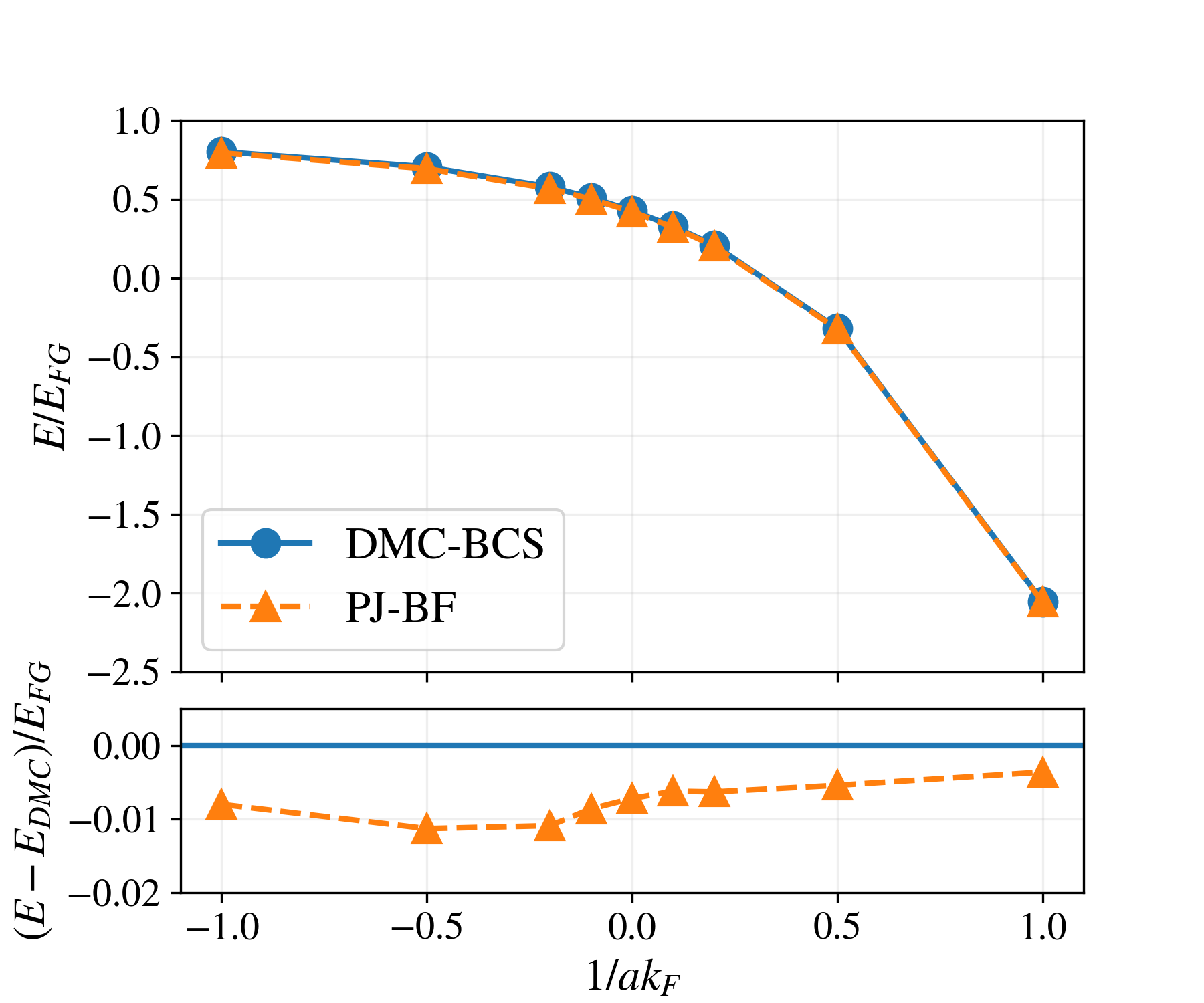}
  \caption{Upper panel: Energy per particle in the BCS-BEC crossover region as a function of the scattering length $a$ for a fixed effective range $k_F r_e = 0.2$. Lower panel: Difference between Pfaffian-Jastrow with backflow (PJ-BF) and DMC-BCS benchmark energies. See Table~\ref{tab:int-params} for the corresponding values of $v_0$ and $\mu$. }
\label{fig:E-akF}
\end{figure}

Finally, we explore the BCS-BEC crossover region for a fixed effective range $k_F r_e = 0.2$ in Fig.~\ref{fig:E-akF}. See Table~\ref{tab:int-params} for the values of the interaction parameters $v_0$ and $\mu$, as well as the corresponding DMC-BCS benchmarks and the PJ ansatz results. The cases closer to unitarity were used to pretrain the cases further away. In the BCS regime, our PJ ansatz consistently yields energies $\sim 0.01 E_{FG}$ lower than those obtained from DMC-BCS, albeit with slightly inferior performance in the BEC regime. We attribute this effect to the need for increased flexibility in capturing the short-range behavior of pairs in the BEC regime. Simply increasing the size of feedforward neural network $\nu$ that defines the pairing orbital should alleviate this discrepancy. In any case, the PJ ansatz gives lower energies than DMC-BCS for all scattering lengths tested.

\begin{table}[t]
  \centering
    \begin{tabular}{ |c||c|c||c|c| } 
    \hline
    $1/ak_F$ & $v_0$ & $\mu$ & DMC-BCS & PJ-BF \\
    \hline\hline
    -1 & 0.879214 & 11.06247 & 0.801(1) & 0.7930(2)\\
    -0.5 &  0.933216 & 10.55715 &  0.705(1) & 0.6937(3)\\
    -0.2 & 0.971423 & 10.23012 & 0.578(1) & 0.5671(3)\\
    -0.1 & 0.985366 & 10.11637 & 0.510(1) & 0.5014(4)\\
    0 & 1.0 & 10.0 & 0.428(1) & 0.4208(3)\\
    0.1 & 1.015388 & 9.880801 & 0.328(1) & 0.3218(3)\\
    0.2 & 1.031602 & 9.758564 & 0.208(1) & 0.2017(3)\\
    0.5 & 1.086081 & 9.371025 & -0.319(1) & -0.3244(4) \\
    1 & 1.204354 & 8.632898 & -2.053(1) & -2.0566(6) \\
    \hline
    \end{tabular}
\caption{Energies per particle and parameters for the two-body potential in Eq.~\eqref{eq:int} giving different scattering lengths with the same effective range $k_F r_e = 0.2$.}
\label{tab:int-params}
\end{table}

\section{Conclusions and perspectives}
\label{sec:conclusion}
In this study, we propose a novel neural-network quantum state based on the Pfaffian-Jastrow (PJ) framework that utilizes a message-passing neural network (MPNN) to encode pairing and backflow (BF) correlations. We evaluate its performance against comparable Slater-Jastrow (SJ) ans\"atze with identical MPNN architectures. Our results indicate that increasing the depth of the MPNN systematically improves the performance of the SJ ans\"atze, but backflow correlations within the single-particle picture are still insufficient in capturing all pairing correlations. However, we demonstrate that a simple and compact PJ-BF ansatz surpasses the DMC-BCS benchmark with ease. 

Transfer learning has proven to be an essential tool in this work. It enables the realization of the unitary limit in a controlled manner, mitigating the risk of becoming trapped in local minima. It also allows for the efficient exploration of regions beyond unitarity, unlocking new avenues for studying the BCS-BEC crossover. 
Transfer learning will remain a crucial part of our training procedure as we move to larger systems. All unpolarized systems can be treated with a single architecture, while the $N\pm 1$ systems can be treated by introducing one additional FNN to represent the unpaired single-particle orbital. This modification is straightforward to implement, making the calculation of the gap more accessible and enabling further advancements in our work. 

Besides calculating the gap, our next steps include a direct comparison with the STU Pfaffian wave function of Ref.~\cite{Bajdich:2008}. We also plan to perform a more careful extrapolation to the $r_e \rightarrow 0$ limit since we have used relatively large values of $k_Fr_e$ for this initial investigation. However, more hyperparameter tuning will be needed, especially about the width of the hidden layers, since the smaller values of $r_e$ will require more flexibility. 

Our Pfaffian-Jastrow-Backflow NQS displays immense potential in the study of ultra-cold Fermi gases. Unlike conventional methods, our PJ-BF ansatz is not subject to biases arising from physical intuition or a lack thereof, as it does not require specifying a particular form for the pairing orbitals. For this reason, it can be readily applied to other strongly-correlated systems, including molecules and other strongly-correlated quantum systems. In stark contrast to the commonly used geminal wave function, our ansatz does not rely on ordering the spin of the interacting fermions, and it is therefore amenable to Hamiltonians that exchange spin, such as those modeling nuclear dynamics. In this regard, we anticipate calculations of atomic nuclei and low-density isospin-asymmetric nucleonic matter and carry out detailed investigations on the nature of nuclear pairing~\cite{Dean:2002zx}.  

When the stochastic reconfiguration algorithm and transfer learning techniques are combined with the enforcement of translational, parity, and time-reversal symmetries, highly non-perturbative correlations can be encoded in a small number of parameters by modern standards. 
This approach will pave the way for future developments in the study of many-body systems, as it offers a powerful tool for encoding correlations in a compact and computationally feasible manner.

\vspace{0.1cm}

Note Added: A work very recently appeared in pre-print~\cite{lou2023neural} introduces neural backflow transformations in a geminal wave function and studies the unitary Fermi gas. We leave systematic comparisons between the two approaches to future works while already observing that the Pfaffian wave function is a strict generalization of the Geminal state~\cite{Genovese:2020, Bajdich:2008}.

\section{Acknowledgments}

The work of JK and MHJ is supported by the U.S. National Science Foundation Grants No. PHY-1404159 and PHY-2013047. AL and BF are supported by the U.S. Department of Energy, Office of Science, Office of Nuclear Physics, under contracts DE-AC02-06CH11357, by the 2020 DOE Early Career Award number ANL PRJ1008597, by the NUCLEI SciDAC program, and Argonne LDRD awards.
The work of SG is supported by the U.S. Department of Energy, Office of Science, Office of Nuclear Physics, under contract No.~DE-AC52-06NA25396, by the U.S. Department of Energy, Office of Science, Office of Advanced Scientific Computing Research, Scientific Discovery through Advanced Computing (SciDAC) NUCLEI program, and by the Department of Energy Early Career Award Program. 
The work of GP, JN, GC is supported by the Swiss National Science Foundation under Grant No. 200021\_200336, and by Microsoft Research. 
Computer time was provided by the Los Alamos National Laboratory Institutional Computing Program, which is supported by the U.S. Department of Energy National Nuclear Security Administration under Contract No. 89233218CNA000001.

\bibliography{references}

\appendix
\section{Backflow Orbitals}
\label{appx:bf}

To motivate the form of the backflow transformation in Eqs.~\eqref{eq:bf-r} and \eqref{eq:bf-chi}, let us first revisit the original plane wave orbitals in Eq.~\eqref{eq:pw-phi}. We simulate our system in the basis 
\begin{equation}
| \bm{x}_i \rangle
= 
| \bm{r}_i \rangle
| s_i \rangle,
\end{equation}
where $| \bm{r}_i \rangle$ are eigenstates of the position operator, with $\bm{r}_i \in \mathbb{R}^d$, and $|s_i\rangle$ are eigenspinors of the $S_z$ operator, with $s_i\in \{\uparrow, \downarrow\}$. In the fixed-node approximation, we take the single-particle states to be products of momentum eigenstates with definite wave vector $\bm{k}_\alpha = 2 \pi \bm{n}_\alpha/L$, $\bm{n}_\alpha\in \mathbb{Z}^d$, and eigenspinors with definite spin projection $s_\alpha$,
\begin{equation}
| \phi_\alpha \rangle = | \bm{k}_\alpha \rangle | s_\alpha \rangle.
\end{equation}
Omitting overall normalization constants, the probability amplitude of measuring particle $i$ in state $\alpha$ is
\begin{align}
\phi_\alpha(\bm{x}_i) 
= \langle \bm{x}_i | \phi_\alpha \rangle
= \langle \bm{r}_i | \bm{k}_\alpha \rangle \langle s_i  | s_\alpha \rangle = e^{i \bm{k}_\alpha \cdot \bm{r}_i} \delta_{s_\alpha, s_i},
\end{align}
which we call the plane wave orbitals.

Now, let us transform to a new basis with modified position eigenstates and a superposition of eigenspinors
\begin{equation}
|\tilde{\bm{x}}_i \rangle 
=
|\tilde{\bm{r}}_i \rangle 
|\chi_i \rangle. 
\end{equation}
While permutation equivariance is the sole essential property required for the backflow transformation $|\bm{x}_i \rangle \mapsto |\tilde{\bm{x}}_i \rangle$ to preserve the antisymmetry of the fermionic wave function, an additional property is desirable for computational convenience. Specifically, when the transformation depends on certain parameters, we aim to have $|\tilde{\bm{x}}_i \rangle = |\bm{x}_i \rangle $ when the parameters are identically zero. Then, nonzero parameters signify deviations from the original plane wave orbitals, such that less training is required compared to completely trainable orbitals. 

An appropriate spatial transformation is trivial. We simply define new parameters $\bm{u}_i\in\mathbb{C}^d$, called the backflow displacement, and shift the coordinates as $\bm{r}_i' = \bm{r}_i + \bm{u}_i$.
The backflow displacement is complex, allowing for changes in both the phases and amplitudes of the original plane wave orbitals. 

For the spin part of the transformation, we look to spinors on the Bloch sphere for inspiration,
\begin{equation}
| \chi_i \rangle 
=
\cos\left(\frac{\theta_i}{2} \right) | s_i \rangle 
+
\sin\left(\frac{\theta_i}{2} \right) \sigma_i^x | s_i  \rangle.
\label{eq:bloch-spinor}
\end{equation}
In the above, we have introduced another backflow variable $\theta_i \in \mathbb{R}$ akin to the polar angle of a Bloch spinor, and we have excluded the relative phase in favor of a completely real-valued wave function. We also write the superposition in terms of $| s_i \rangle$ and the Pauli $X$-operator $\sigma_i^x$, which flips the spin of the $i$-th particle, rather than $|\uparrow \rangle$ and $|\downarrow \rangle$. This way, it is obvious that $|\chi_i\rangle = | s_i \rangle$ when $\theta_i = 0$, as desired. The overlap of two spinors is given by
\begin{equation}
\begin{split}
\langle \chi_i | \chi_j \rangle
&= \cos \left( \frac{\theta_i}{2} \right) \cos \left( \frac{\theta_j}{2} \right) \langle s_i | s_j \rangle\\
& \hspace{3mm}
+ \sin \left( \frac{\theta_i}{2} \right) \cos \left( \frac{\theta_j}{2} \right) \langle s_i | \sigma_i^{x\dagger} | s_j \rangle \\
& \hspace{3mm} 
+ \cos \left( \frac{\theta_i}{2} \right) \sin \left( \frac{\theta_j}{2} \right) \langle s_i | \sigma_j^x  | s_j \rangle\\
& \hspace{3mm} 
+ \sin \left( \frac{\theta_i}{2} \right) \sin \left( \frac{\theta_j}{2} \right) \langle s_i |  \sigma_i^{x\dagger} \sigma_j^x | s_j \rangle\\
&= \left[ 
\cos \left( \frac{\chi_i}{2} \right) \cos \left( \frac{\chi_j}{2} \right) 
+ \sin \left( \frac{\chi_i}{2} \right) \sin \left( \frac{\chi_j}{2} \right)
\right] \delta_{s_i, s_j}\\
& \hspace{4mm}
\left[
\sin \left( \frac{\theta_i}{2} \right) \cos \left( \frac{\theta_j}{2} \right) 
-\cos \left( \frac{\theta_i}{2} \right) \sin \left( \frac{\theta_j}{2} \right)
\right]
(1-\delta_{s_i, s_j})\\
&= \cos \left( \frac{\theta_i - \theta_j}{2} \right) \delta_{s_i, s_j}
+
\sin \left( \frac{\theta_i - \theta_j}{2} \right) (1-\delta_{s_i, s_j}),
\end{split}
\end{equation}
where the limiting cases are summarized in the following table.
\begin{center}
\begin{tabular}{|c||c|c|}
\hline
$\langle \chi_i | \chi_j \rangle$ & $\theta_i = \theta_j$ & $\theta_i - \theta_j = \pm \pi$\\
\hline\hline
$s_i = s_j$ & 1 & 0\\
\hline
$s_i \neq s_j$ & 0 & $\pm 1$\\
\hline
\end{tabular}
\end{center}

Finally, we can compute the backflow orbitals with the transformed degrees of freedom
\begin{equation}
\begin{split}
\phi_\alpha(\tilde{\bm{x}}_i)
&= \langle \tilde{\bm{x}}_i | \phi_\alpha \rangle\\
&= \langle \tilde{\bm{r}}_i | \bm{k}_\alpha  \rangle
\langle \chi_i | s_\alpha \rangle\\
&= e^{i \bm{k}_\alpha  \cdot \tilde{\bm{r}}_i }
\left( 
\cos \left( \frac{\theta_i}{2} \right) \langle s_i | s_\alpha  \rangle
+ \sin \left( \frac{\theta_i}{2} \right) \langle s_i | \sigma_i^{x\dagger} | s_\alpha \rangle
\right)\\
&= e^{i \bm{k}_\alpha \cdot (\bm{r}_i + \bm{u}_i) } 
\left( 
\cos \left( \frac{\theta_i}{2} \right) \delta_{s_\alpha , s_i}
+ \sin \left( \frac{\theta_i}{2} \right) (1-\delta_{s_\alpha , s_i})
\right).
\end{split}
\end{equation}

In Eqs.~\eqref{eq:bf-r} and \eqref{eq:bf-chi}, we use the notation $\bm{u}_i(X)$ and $\theta_i(X)$ to emphasize that the backflow ``parameters" we define here are not variational parameters, but a function of all other particles. More specifically, they are permutation-equivariant functions of the original $\bm{x}_i$, whose functional forms depend on the outputs of the permutation-equivariant message-passing neural network (MPNN) described in Sec.~\ref{subsubsec:mpnn}. 

\end{document}